\documentclass{aa}  
\usepackage{graphicx}
\usepackage{txfonts}
\usepackage{hyperref}
\usepackage{breakurl} 
\usepackage{etoolbox}
\makeatletter
\patchcmd\@combinedblfloats{\box\@outputbox}{\unvbox\@outputbox}{}{%
   \errmessage{\noexpand\@combinedblfloats could not be patched}%
}%
 \makeatother

\begin{document}

   \title{Spectroscopic EUV observations of impulsive solar energetic particle event sources 
}

%   \subtitle{I. Overviewing the $\kappa$-mechanism}

   \author{R.~Bu\v{c}\'ik\inst{1,}\inst{2} \and A.~Fludra\inst{3} \and R.~G\'omez-Herrero\inst{4} \and D.~E.~Innes\inst{2}  \and B.~Kellett\inst{3}  \and R.~Kumar\inst{5} \and \v{S}.~Mackovjak\inst{6}
          }

   \institute{Institut f\"{u}r Astrophysik, Georg-August-Universit\"{a}t G\"{o}ttingen, D-37077 G\"{o}ttingen, Germany
   \and
   Max-Planck-Institut f\"{u}r Sonnensystemforschung, D-37077 G\"{o}ttingen, Germany, \email{bucik@mps.mpg.de}
   \and
   STFC Rutherford Appleton Laboratory, Didcot, OX11 0QX, UK
   \and 
   Space Research Group, University of Alcal\'a, E-28871 Alcal\'a de Henares, Spain
   \and 
   Department of Astrophysical Sciences, Princeton University, Princeton, NJ 08544, USA 
   \and
   Institute of Experimental Physics, Slovak Academy of Sciences, Watsonova 47, Ko\v{s}ice, Slovakia
             }

   %\date{Received September 15, 1996; accepted March 16, 1997}
   \date{Received ; accepted }

% \abstract{}{}{}{}{} 
% 5 {} token are mandatory
 
  \abstract
   {The remote observations of solar flare ion acceleration are rather limited. There are theoretical predictions for signatures of ion acceleration in EUV line profiles. Previous tests involve observations of flares with no evidence for energetic ions.}
   {We aim to examine a source flare of impulsive (or \element[ ][3]{He}-rich) solar energetic particle events with EUV line spectroscopy.}
   {We inspect all (90$+$) reported \element[ ][3]{He}-rich flares of previous solar cycle 23 and find only four (recurrent) jets in the field of view of SOHO CDS. The jet with the most suitable spatial and temporal coverage is analyzed in detail. }
   {Two enhanced (non-thermal) line broadenings are observed in the cooler chromospheric / transition-region lines and they are localized near the site where the closed magnetic loops reconnect with the open magnetic field lines. Both enhanced broadenings are found in the sites with redshifts in the lines, surrounded by the region with blueshifts. One enhanced line broadening is associated with a small flare without energetic particle signatures while another occurs just after the particle acceleration signatures of the main flare terminated. 
}
   {The observed excess broadening appears to be not directly related to the energetic ion production and motions. Further investigations where the critical impulsive phase of the flare is covered are required, ideally with high-resolution spectrometers intentionally pointed to the \element[ ][3]{He}-rich solar energetic particle source.}

               
   \titlerunning{Spectroscopic EUV observations of impulsive SEP event sources}
   \maketitle
%
%-------------------------------------------------------------------

\section{Introduction}

Impulsive (or \element[ ][3]{He}-rich) solar energetic particle (SEP) events are characterized by a peculiar ion composition, markedly different from the corona or solar wind \citep[][and references therein]{mas07}. The heavy ion enhancement in the impulsive SEP events is an increasing function of ion mass. For example, \element[ ][ ]{Fe} is enhanced by a factor of about 10 compared to its coronal abundance; while the enrichment of ultra-heavy (UH) ions readily exceed an enhancement factor of 100. This pattern conflicts with the enhancement of low-mass \element[ ][3]{He} that is enhanced by factors up to 10$^4$. It is believed that the anomalous composition is the signature of a unique acceleration process operating at the flare sites. A resonant interaction with plasma waves has been commonly involved in models of ion acceleration and fractionation in solar flares \citep{fis78,win89,tem92,mil98,zha04,liu06,kum17} though another, non-wave, mechanisms (scattering on magnetic islands or DC electric field) have been suggested \citep[e.g.,][]{dra09,fle13}. Electrons are almost always accelerated to sub-relativistic speeds in these events as demonstrated by direct in-situ measurements \citep{rea85,wan12} or by detection of type III radio bursts \citep{rea86,nit06}.

\element[ ][3]{He}-rich SEP events have been associated with minor (mostly B- and C-class) soft X-ray flares \citep{rea88a,rea04,nit06,nit15,buc14,buc16}. In addition, an inverse correlation between the soft X-ray peak intensity and \element[ ][3]{He} (UH) enrichment has been reported \citep{rea88a,rea04}. \element[ ][3]{He}-rich flares have been commonly observed as a jet-like form in extreme ultraviolet \citep[EUV;][]{nit06,nit15,che15,buc18} or white-light coronograph images \citep{kah01,wan06}. The association with jets has been interpreted as an evidence for magnetic reconnection involving field lines open to interplanetary space \citep{shi92}. Remote observations of ion acceleration in solar flares are quite limited. The energetic ions do not show clear flare signatures as energetic electrons do through the hard X-ray (HXR) bremsstrahlung or radio emissions. Gamma-ray nuclear lines, produced in a bombardment of energetic ($>$1--10\,MeV\,nucleon$^{-1}$) ions with ambient nuclei in the chromosphere, were observed only in major (X- or M-class) flares \citep[e.g.,][]{vil11}. Although the abundances derived from $\gamma$-ray lines show \element[ ][3]{He} and heavy-ion enrichment \citep{mur91,man99}, the $\gamma$-ray lines were generally not observed in \element[ ][3]{He}-rich SEP events \citep[but see][for the exceptions]{hol90,kar07}. 
 
There are theoretical predictions for the formation of asymmetrical broadened profiles (extended red wings) in \ion{H}{I} Lyman-$\alpha$ (1216\,$\AA$) and \ion{He}{II} Lyman-$\alpha$ (304\,$\AA$) atomic lines \citep[e.g.,][]{orr76,can85,pet90,bro99} caused by the accelerated ions in solar flares. Specifically, the low-energy ($<$1\,MeV\,nucleon$^{-1}$) protons or $\alpha$-particles (invisible in $\gamma$-ray lines) can capture electrons due to charge-exchange with thermal neutral hydrogen in the chromosphere and radiate non-thermal Doppler-broadened Lyman-$\alpha$. The profiles are expected to extend rather far into red wings, $\sim$10\,$\AA$ for \ion{H}{I} and 4--5\,$\AA$ for \ion{He}{II}. Similar processes of charge-exchange between suprathermal (keV) protons and residual neutral hydrogen in the corona have been discussed by \citet{lam13}. The authors have suggested that the suprathermal ions should manifest themselves through extended wings in resonantly scattered chromospheric \ion{H}{I} Lyman-$\alpha$ in the corona. However, so far, no clear evidence of Doppler-broadened Lyman-$\alpha$ from energetic ions in solar flares has been observed \citep{bro01,hud12} though the effect has been reported in a flare on the red dwarf star AU Mic \citep{woo92}. \citet{bro01} has examined C-class flare with a SOHO CDS spectrometer for a signal of enhanced red wings in the profile of \ion{He}{II} Ly$\alpha$ line. \citet{hud12} have examined nine hard X-ray/$\gamma$-ray events (in M- or X-class flares) for signatures of broad wings in \ion{He}{II} Ly$\alpha$ using the full-Sun EVE instrument aboard SDO. The authors discuss that strong chromospheric heating during the flare may reduce or mask the charge-exchange mechanism. \citet{jef16,jef17} have analyzed the \ion{Fe}{XVI} 262.98\,$\AA$ and \ion{Fe}{XXIII} 263.76\,$\AA$ line profiles observed by the Hinode EIS in two X-class flares and found the profiles consistent with a kappa distribution of emitting ions with kappa values corresponding to the highly accelerated ion distribution. The authors have speculated that a flare accelerated Fe ions could account for the observed excess broadening.
 
A suitable candidate for EUV spectral line investigations, with a low background, may be \element[ ][3]{He}-rich SEP events whose solar sources have been associated with minor X-ray flares or weak EUV brightenings. In addition, in such events, we have direct evidence for ion acceleration in the flare via in-situ measurements of energetic ion fluxes. In this study, we present the first-time observations of EUV spectral lines in the source flare of \element[ ][3]{He}-rich SEPs.
 
%--------------------------------------------------------------------
\section{Methods}

We study the solar source of \element[ ][3]{He}-rich SEPs with the Coronal Diagnostic Spectrometer \citep[CDS;][]{har95} aboard Solar and Heliospheric Observatory (SOHO). The CDS consists of normal incidence (NIS) and grazing incidence (GIS) spectrometers. The observations examined in this study were obtained by NIS using a 4$\arcsec$ $\times$ 240$\arcsec$ slit to produce a raster image. The NIS covers 307--379\,$\AA$ and 513--633\,$\AA$ wavelength ranges, with approximate spectral resolutions 0.32\,$\AA$ and 0.54\,$\AA$ full width at half-maximum (FWHM), respectively. The following three lines are examined: \ion{He}{I} 584.30\,$\AA$, \ion{O}{V} 629.39\,$\AA$, and \ion{Fe}{XVI} 360.80\,$\AA$, corresponding to the formation temperatures $\sim$0.03\,MK, 0.25\,MK, and 2.5\,MK, respectively. These CDS lines have been analyzed in several previous studies on solar flares  \citep[e.g.,][]{cza99,mil06a,mil06b,ter06}. We remind that the  \ion{He}{I} stands for neutral helium, \ion{O}{V} is 4$\times$ ionized oxygen with charge-to-mass ratio $Q/A=$4/16 and \ion{Fe}{XVI} is 15$\times$ ionized iron with $Q/A=$15/56. We also examine EUV images of the \element[ ][3]{He}-rich solar source obtained by SOHO EIT \citep{del95} and the Transition Region and Coronal Explorer \citep[TRACE;][]{han99}, and magnetograms obtained by SOHO MDI \citep{sch95}. The \element[ ][3]{He}-rich SEPs are analyzed using the measurements from SOHO Electron Proton Helium Instrument \citep[EPHIN;][]{mul95}. The EPHIN, a $dE/dx$ versus $E$ telescope, measures hydrogen and helium isotopes in the range 4\,MeV\,nucleon$^{-1}$ to $>$53\,MeV\,nucleon$^{-1}$. We also inspect dynamic radio spectra for the event associated type III radio bursts. The radio data are provided by the Wind WAVES instrument \citep{bou95} with a frequency range ($<$16\,MHz) covering emission generated from about 2\,$R_{\sun}$ to 1\,au and Phoenix-2 (Bleien, Switzerland) broadband spectrometer \citep{mes99} operated in the frequency range 0.1--4\,GHz. Furthermore, we make use of soft and HXR observations from NOAA GOES-10 XRS and Wind KONUS \citep{apt95} sensors, respectively. The interplanetary magnetic field (IMF) and solar wind speed measurements were provided by the MAG \citep{smi98} and SWEPAM \citep{mcc98} instruments on ACE. Note that while SOHO, ACE and Wind were in the interplanetary space the TRACE was in geocentric orbit. 

%----------------------------------------------------------------- 
\begin{figure}
%   \begin{figure}
   \centering
     \includegraphics[width=0.46\textwidth]{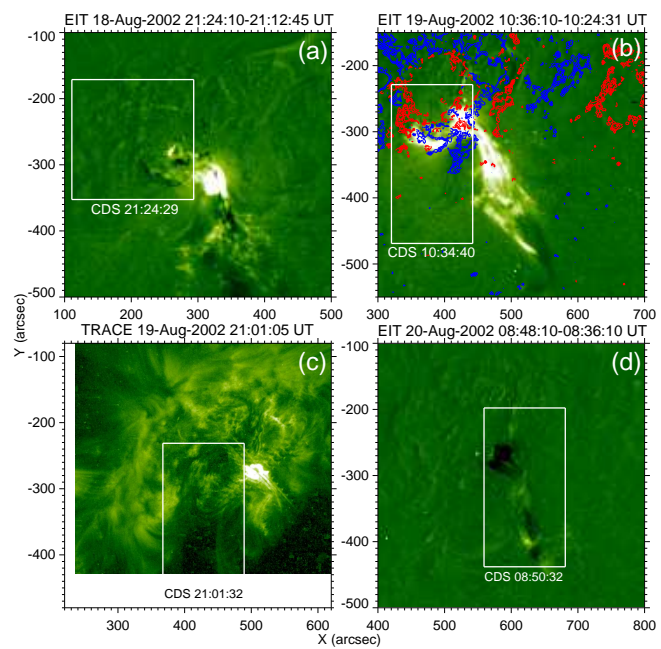}
      \caption{SOHO EIT 195\,$\AA$ EUV 12-min running differences images for flares in event 1 (panel a), event 2 (panel b) and event 4 (panel d). Over-plotted in panel (b) are $\pm$500\,G SOHO MDI magnetogram data contours for 2002 August 19 11:11\,UT, where blue/red is positive/negative polarity. TRACE 195\,$\AA$ EUV direct image (re-scaled to L1 and aligned with CDS \ion{Fe}{XVI} 360.80\,$\AA$ line raster image) for event 3 where SOHO EIT data are not available (panel c). Note that empty area at bottom of the panel (c) is due to a restricted TRACE FOV. White rectangles depict the CDS FOV. While the images in events 1, 2, 3 are near ($\pm$1--2 minutes) the flare maximum the image in event 4 is $\sim$20 minutes after the flare maximum. 
              }
         \label{fig01}
%   \end{figure}
   \end{figure}
%-----------------------------------------------------------------

To find a \element[ ][3]{He}-rich solar source in the CDS field-of-view (FOV), we inspect all (93) \element[ ][3]{He}-rich SEP events of previous solar cycle 23 with a reported solar source \citep{mas02,nit06,wan06}. The solar sources in 67 events have been identified by \citet{nit06}, in another 19 events by \citet{wan06} and in 7 events by \citet{mas02}. These events are from the period September 1997 to March 2003. Note that the CDS started scientific operation in April 1996. To identify the solar source of a \element[ ][3]{He}-rich SEP event, \citet{wan06} examined H$\alpha$ and SOHO EIT flares within a few hours of the estimated ion injection time; \citet{nit06} examined type III radio bursts in the 5-hour window preceding the observed ion onset and then searched for the associated brightening in SOHO EIT coronal images; and \citet{mas02} selected the event associated X-ray flare based on the estimated ion injection time. Note that overlays of CDS FOV with other imaging instruments on SOHO are subject to a 5$\arcsec$ pointing uncertainty of the CDS \citep{flu01}.

We found only four \element[ ][3]{He}-rich SEP source flares fully or partially covered by the CDS FOV. All these flares were from the same active region, marked by NOAA number 10069. Fig.~\ref{fig01} shows SOHO EIT or TRACE EUV images of these \element[ ][3]{He}-rich SEP sources with superimposed CDS FOVs. Fig.~\ref{fig01}a and Fig.~\ref{fig01}c indicate that the flares in event 1 and 3 were only partially captured with CDS. The flare in event 2 was quite well captured (see Fig.~\ref{fig01}b) and therefore is further investigated in detail. In event 4 only the post-flare phase was captured (see Fig.~\ref{fig01}d). Note that a dimming jet exhibited an interesting twisted configuration in event 4. Table~\ref{tab1} shows some characteristics of these flares and the associated particle events. Column 1 gives the event number, column 2 the particle event start time, and column 3 the type III radio burst start time. Columns 4, 5, 6, and 7 indicate the GOES X-ray flare start time, time at the flare maximum, the flare location, and class, respectively. Columns 8 and 9 provide the \element[ ][3]{He}/\element[ ][4]{He} and Fe/O abundance ratios, respectively. Table~\ref{tab2} lists some parameters of CDS studies that match these \element[ ][3]{He}-rich flares. Column 1 gives the event number, column 2 the start time of the raster image, column 3 the duration of the rastering, column 4 the temporal resolution, and column 5 the exposure time. Columns 6, 7, and 8 indicate the CDS pointing, FOV, and the spatial resolution, respectively. Columns 9 and 10 indicate the names of the CDS studies and their total durations, respectively. Note that the observational mode in event 2 includes four lines \ion{He}{I}, \ion{O}{V}, \ion{Mg}{IX}, and \ion{Fe}{XVI}.

\begin{table*}
\tabcolsep 5.pt
%\begin{sidewaystable*}
\caption{\element[ ][3]{He}-rich SEP event properties}
\label{tab1}
\centering
\begin{tabular}{c c c c c c c c c c}
\hline\hline
Event & SEP start\tablefootmark{a} & Type III & \multicolumn{4}{c}{1--8\,$\AA$ GOES X-ray flare\tablefootmark{c}} & \element[ ][3]{He}/\element[ ][4]{He}\tablefootmark{d} & Fe/O\tablefootmark{e} & Ref.\\
\cline{4-7}
& & start\tablefootmark{b} & Start & Max & Location & Class & & \\
\hline
1 & 2002-Aug-18 23:00 & 21:10 & 21:12 & 21:25 & S13W20 & M2.2 & 0.02$\pm$0.01 & 0.40$\pm$0.04 & 1,2,3\\
2 & 2002-Aug-19 11:00 & 10:30 & 10:28 & 10:34 & S12W26 & M2.1 & 0.42$\pm$0.04 & 1.34$\pm$0.04 & 1,2,3,4,5,6,7\\
3 & 2002-Aug-19 23:00 & 20:57 & 20:56 & 21:02 & S11W32 & M3.1 & 0.43$\pm$0.08 & 2.12$\pm$0.07 & 1,2,7,8\\
4 & 2002-Aug-20 10:00 & 08:25 & 08:22 & 08:26 & S11W38 & M3.4 & 0.06$\pm$0.02 & 1.96$\pm$0.04 & 1,2,3,4,7,8,9\\
\hline
\end{tabular}
\tablefoot{
\tablefoottext{a}{Wind LEMT 2--3\,MeV\,nucleon$^{-1}$ \citep{nit06}}
\tablefoottext{b}{Wind WAVES 4\,kHz--13.8\,MHz \citep{nit06}}
\tablefoottext{c}{from \citet{nit06} but start time (Start) and time at flux maximum (Max) from Solar and Geophysical Event Reports compiled by the NOAA Space Weather Prediction Center (SWPC); \url{ftp://ftp.swpc.noaa.gov/pub/warehouse/2002/}}
\tablefoottext{d}{SOHO EPHIN 5--25\,MeV\,nucleon$^{-1}$ (this work)}
\tablefoottext{e}{Wind LEMT 3.3--10\,MeV\,nucleon$^{-1}$ \citep{rea04}}
}
\tablebib{
(1) \citet{nit06}; (2) \citet{rea04}; (3) \citet{can10}; (4) \citet{les03}; (5) \citet{wan06}; (6) \citet{tan08}; (7) \citet{rea14}; (8) \citet{kru11}; (9) \citet{wie10}
}
\end{table*}
%\end{sidewaystable*}
%--------------------------------------------------------------------------------

\begin{table*}
\tabcolsep 4.pt
%\small
%\begin{sidewaystable*}
\caption{SOHO CDS data for \element[ ][3]{He}-rich SEP sources}
\label{tab2}
\centering
\begin{tabular}{c c c c c c c c c c}
\hline\hline
SEP & Raster start time & Dur. & Tem. res. & Exp. time & Pointing & FOV & Spatial res. &  \multicolumn{2}{c}{CDS study\tablefootmark{a}}\\
\cline{9-10}
event & & (sec) & (sec) & (sec) & (arcsec) & (arcsec) & (arcsec) & Name & Dur. \\
\hline
1 & 2002-Aug-18 21:24:29 & 681 & 15.1 & 8 & 202.26, -261.81 & 182.80$\times$181.40 & 4.06$\times$3.36 & FLARE\_AR & 8.4\\
2 & 2002-Aug-19 10:34:40 & 318 & 10.6 & 5 & 381.22, -348.87 & 121.90$\times$240.20 & 4.06$\times$1.68 & CD5 & 4.8\\
3 & 2002-Aug-19 21:01:32 & 321 & 10.7 & 5 & 428.32, -351.31 & 121.90$\times$240.20 & 4.06$\times$1.68 & CD5 & 10.1\\
4 & 2002-Aug-20 08:50:32 & 318 & 10.6 & 5 & 620.10, -317.88 & 121.90$\times$240.20 & 4.06$\times$1.68 & CD5 & 2.2\\
\hline
\end{tabular}
\tablefoot{
\tablefoottext{a}{FLARE\_AR: Flaring Active Region Study, CD5: Coronal Dynamics; Dur. (hr)}
}
\end{table*}
%\end{sidewaystable*}

%--------------------------------------------------------------------
\section{Results}

Events 1--4 have been included in previous studies (see references in Table~\ref{tab1}). All these events were associated with M-class flares. The events exhibited large ion intensities at high energies ($>$10\,MeV\,nucleon$^{-1}$) and therefore called large impulsive SEP events \citep{rea04} as opposed to more typical small impulsive events. The events are electron-rich (or proton-poor) as reported in a study of \citet{can10}. The events showed large enhancements of trans-Fe elements by a factor of 10$^2$--10$^3$ for $Z>$50 \citep{rea04}. The measured abundances in events 2--4 are consistent with a coronal temperature 2.5--3\,MK in their source \citep{rea14}. \citet{wie10} have performed a detailed elemental and isotopic compositional analysis of the measurements above 10\,MeV\,nucleon$^{-1}$ in event 4 (2002 August 20). Based on the inferred charge states (assuming $Q/A$-dependent fractionation) a best-fit temperature in source plasma was found to be $\sim$4\,MK with large deviations to higher temperatures for Fe--Ni. The authors also reported a temperature $\sim$1.6\,MK, assuming a possible $Q^2/A$ law. They also reported observed Fe charge states $Q$ between 18 (\ion{Fe}{XIX}) and 22 (\ion{Fe}{XXIII}) at low energies (0.25--1\,MeV\,nucleon$^{-1}$). The 2002 August 20 event was the largest impulsive event in solar cycle 23 observed at energies $>$10\,MeV\,nucleon$^{-1}$ \citep{les03}. 

%----------------------------------------------------------------- 
\begin{figure}
%   \begin{figure}
   \centering
      \includegraphics[trim= {0cm 3cm 0cm 0cm}, clip,width=0.45\textwidth]{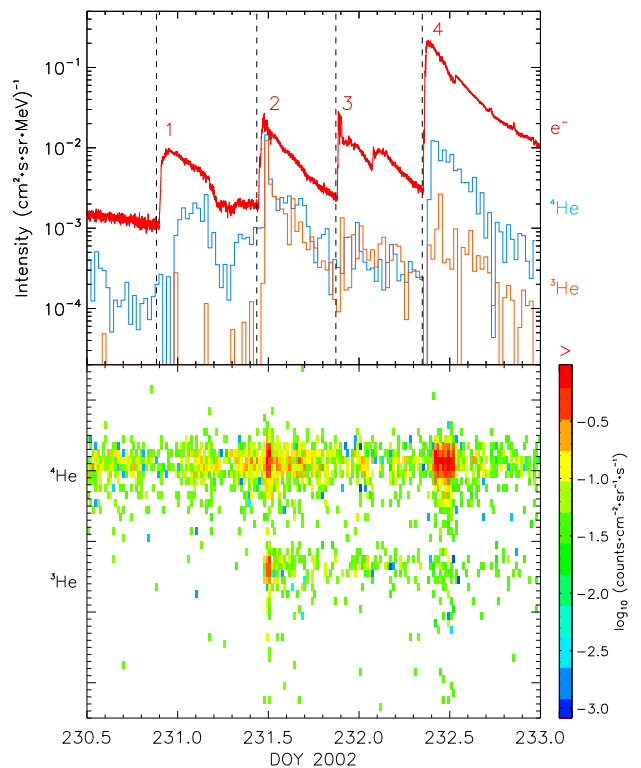}
      \caption{SOHO EPHIN measurements. Top: 1-min averaged 0.7--3.0\,MeV electron intensity (divided by 500) and 30-min averaged 10--25\,MeV\,nucleon$^{-1}$ \element[ ][3]{He}, \element[ ][4]{He} intensities. Dashed vertical lines indicate X-ray flares associated with events 1--4 (see Table~\ref{tab1}). Bottom: helium mass spectrogram of individual PHA ions with no energy restriction. 
              }
         \label{fig02}
%   \end{figure}
   \end{figure}
%-----------------------------------------------------------------

The EUV flare type in the investigated event 2 has been noted as an ejection in 195\,$\AA$ SOHO EIT images in \citet{wan06}. Examining the TRACE 195\,$\AA$ images, the flare type in all four events 1--4 has been noted as a jet in \citet{nit06}. The SOHO EIT flare in event 2 occurred close to a new emergence of a negative-polarity flux observed with SOHO MDI magnetograms between 2002 August 19 01:36--11:25\,UT \citep{tan08}. Using the Nan\c{c}ay Radio-Heliograph observations at 164\,MHz, \citet{tan08} found that the metric type III radio bursts were located near the top of the loops at height $\sim$0.15\,$R_{\sun}$. Events 3 and 4 have been included in a study on electron acceleration in solar jets \citep{kru11}. A HXR imaging with RHESSI showed three chromospheric (HXR) sources in event 4, two at the loop foot-points and one related to the open field line. In event 3 only two chromospheric sources were seen, where a third source could be hidden by the extended flare \citep{kru11}. The authors found the onset of the EUV jets observed with TRACE coincident with the HXR emission. For event 2 no RHESSI solar data are available for the associated flare at $\sim$10:07--10:41\,UT on 2002 August 19.

The PFSS (potential-field source-surface) extrapolations of the photospheric magnetic field (National Solar Observatory synoptic maps) in event 2 show the Earth-directed open field lines near the associated flare \citep{wan06}. Specifically, the field lines that cross the source surface at longitudes between W35 and W65 and at latitudes less than 20$\degr$ from the ecliptic (but not those directed to the ecliptic) were rooted next ($<$4$\degr$) to the event source. \citet{wan06} have noticed that a dipole field line that intersects the source surface at 20$\degr$ would continue to bend equatorward beyond the source surface. The PFSS extrapolations based on the MDI magnetograms assimilated into the evolving surface flux model show no open field lines within 10$\degr$ from the flare location \citep{nit06}. Using the Wilcox Solar Observatory synoptic maps the authors report open field lines with a combined offset at the photosphere (from the associated flare) and at the source surface (from the Parker spiral footpoint) of less than 10$\degr$. It has been pointed out that the PFSS extrapolated field depends to some extent on the input magnetograms \citep{nit06}.

\subsection{\element[ ][3]{He}-rich SEP events 1--4}

%----------------------------------------------------------------- 
\begin{figure*}
 %  \begin{figure}
   \centering
     \includegraphics[angle=90, width=0.89\textwidth]{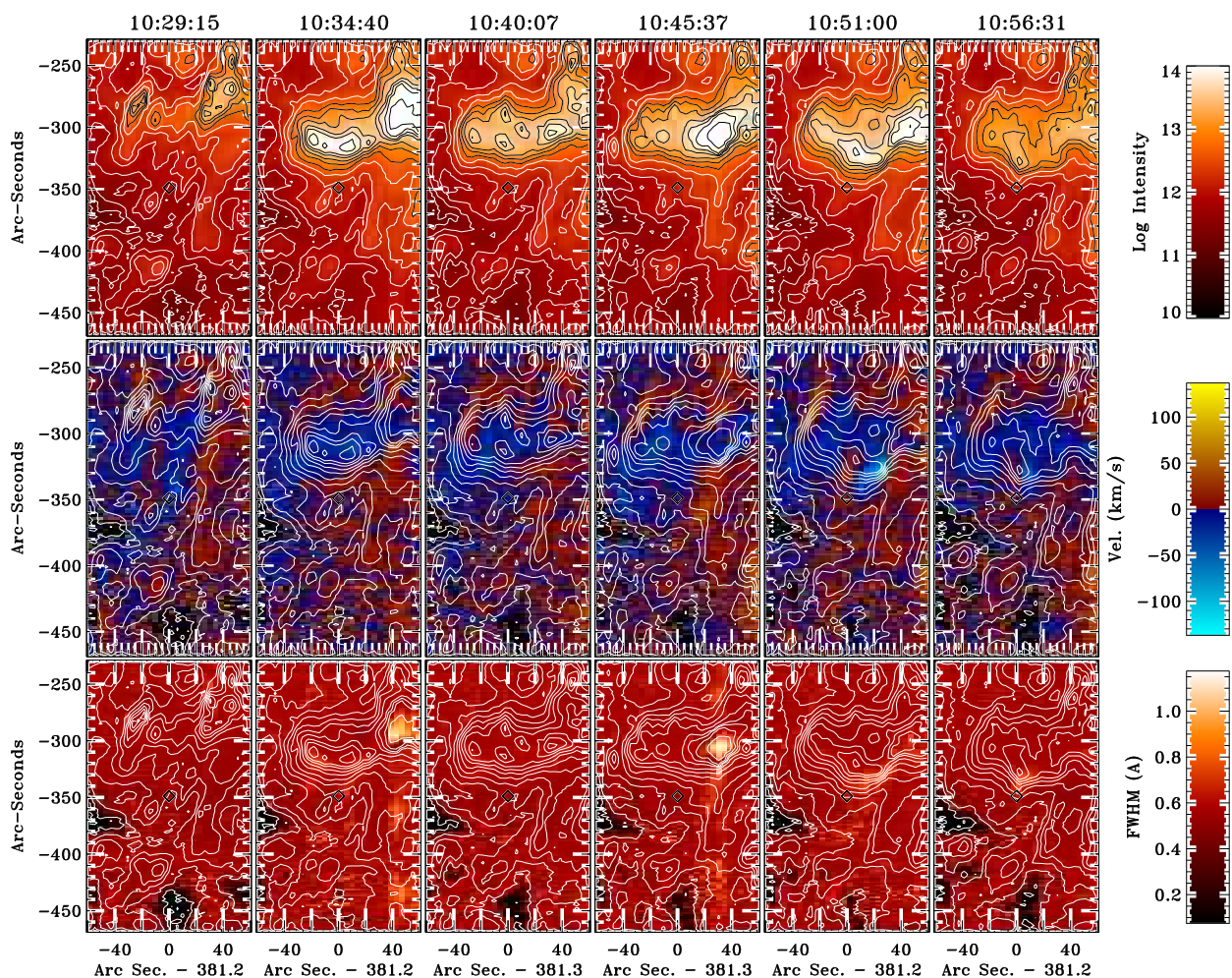}
      \caption{The CDS rasters (\ion{O}{V} 629.39\,$\AA$ line) of the source flare in event 2 between 10:29:15 and 10:56:31\,UT on 2002 August 19. Times are given for the beginning of the rastering. Top row: intensity (photons\;s$^{-1}$\,cm$^{-2}$\,sr$^{-1}$). Middle row: wavelength shift relative to the median wavelength taken from the full image. Bottom row: the FWHM of the line. Over-plotted are the contours for every 0.25 of log$_{10}$ intensity.
              }
         \label{fig03}
 %  \end{figure}
   \end{figure*}
%-----------------------------------------------------------------

Due to a limited element resolution of the Wind LEMT instrument, only an upper limit of 0.2 has been obtained for the \element[ ][3]{He}/\element[ ][4]{He} ratio in events 1--4 at 2.1--2.5\,MeV\,nucleon$^{-1}$ \citep{rea04}. Fig.~\ref{fig02} shows SOHO EPHIN measurements of \element[ ][3]{He}-rich SEP events 1--4 in the period 2002 August 18 (day 230) 12:00\,UT -- August 21 (day 233) 00:00\,UT. The ions with a parallel incidence are selected that improves the isotopic resolution, though at the cost of decreased statistics. Note on the \element[ ][3]{He} increase near the detection limit in event 1. The upper panel of Fig.~\ref{fig02} shows the 0.7--3.0\,MeV electron intensity, and the 10--25\,MeV\,nucleon$^{-1}$ \element[ ][3]{He} and  \element[ ][4]{He} intensities; the lower panel shows the helium mass spectrogram. All four \element[ ][3]{He}-rich SEP events were accompanied by relativistic electron events. The examined event 2 shows the highest  \element[ ][3]{He} intensity and a large \element[ ][3]{He} enrichment. A possible ion event associated with the second electron increase in event 3 is not included in \element[ ][3]{He}/\element[ ][4]{He} in Table~\ref{tab1}.

The electron onset time in event 2 at the energy range 0.25--0.7\,MeV is measured at 10:45\,UT $\pm$1 min. Assuming a scatter-free transport along the nominal Parker spiral (length 1.1\,au for a measured solar wind speed $\sim$500\,km\,s$^{-1}$), the propagation time for electrons with a mean energy 0.42\,MeV (speed 0.84$c$) would be 11 minutes. It implies the solar release time around 10:34\,UT. Subtracting a photon travel time of 8.3\,min, the type III radio bursts start time at the Sun is at $\sim$10:22\,UT in event 2 (see Table~\ref{tab1}). Thus, the inferred release of relativistic electrons is delayed by 12\,min from the type III bursts. A longer travel distance and/or scattering during the transport may be responsible for a delay \citep[e.g.,][]{kah07}. Because of several assumptions, the type III radio burst observations provide a more reliable measurement of the electron release time than the estimate based on electron propagation time. There is no significant proton increase associated with event 2, instead, the event is clear at the helium channels. To obtain an accurate timing for ions is difficult because of low statistics, further complicated by the high particle flux which increases background and triggers the instrument geometric factor into a smaller one at 11:14\,UT on August 19. Using the pulse-height analyzed (PHA) data and selecting \element[ ][3]{He} in a narrow band between 10 and 15\,MeV\,nucleon$^{-1}$, the estimated onset time is 11:35$\pm$10 min. Assuming a mean energy around 12.2\,MeV\,nucleon$^{-1}$ implies a solar release time at 10:38$\pm$10 min, which would be compatible with the electron release time at 10:34\,UT. 

\subsection{Imaging observations of the source in event 2}

Fig.~\ref{fig03} shows the 122$\arcsec$ $\times$ 240$\arcsec$ CDS raster images in \ion{O}{V} 629.39\,$\AA$ line taken during different phases of the \element[ ][3]{He}-rich flare in event 2. The rasters were built up by scanning the 4$\arcsec$ $\times$ 240$\arcsec$ slit from the solar west to the east over 30 pointing positions in 5 minutes and $\sim$19 seconds. The rasters in the upper panels display the average intensity in the line profile, in middle-row panels the line shift relative to the median wavelength obtained from the whole raster and in the bottom panels the line FWHM. After the SOHO temporary loss in 1998, the CDS spectral line profiles were found to have changed. To obtain the line shift and width from post-recovery data, we perform a single broadened Gaussian least-squares fit to the line intensity profile in every pixel of the raster. W.~T.~Thompson (1999)\footnote{See CDS Software Note No. 53 (\url{http://solar.bnsc.rl.ac.uk/swnotes/cds_swnote_53.pdf})} reports a broadened Gaussian function that matches the post-recovery line profiles well. The broadened Gaussian function has two extra parameters to fit the asymmetry wings of the CDS instrumental response. The CDS intensity maps in the upper row of Fig.~\ref{fig03} indicate two distinct brightenings, one at 10:34:40\,UT and another at 10:45:37\,UT. These intensity brightenings contain small areas of the enhanced line broadenings as seen in the bottom row of Fig.~\ref{fig03}. The middle-row panels of Fig.~\ref{fig03} indicate that both intensity brightenings roughly correspond to the blueshift (or plasma up-flow) region, though the enhanced line broadenings contain the redshift (down flow) areas or areas with the reduced up-flow velocity (dark blue pixels). The strongest blue shift is seen at 10:51:00\,UT, just after the second brightening, and it is not related to the event associated flare (see Fig.~\ref{fig07}a). 

\begin{figure}
%  \begin{figure}
   \centering
          \includegraphics[angle=90, width=0.47\textwidth]{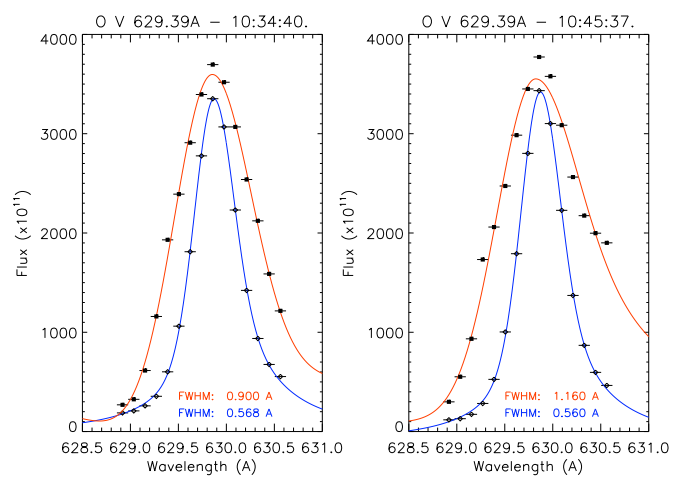}
      \caption{Left: Broadened Gaussian fits to the \ion{O}{V} intensity (photons\;s$^{-1}$\,cm$^{-2}$\,sr$^{-1}$) from the pixel (420$\arcsec$, -290$\arcsec$) (red curve) and to the intensity summed over 90 pixels (to get a similar level of counts) around (370$\arcsec$, -420$\arcsec$) (blue curve) at 10:34:40\,UT on August 19. Right: Similar to the left panel, but at 10:45:37\,UT on August 19, and the red curve is for the pixel (410$\arcsec$, -310$\arcsec$). 
              }
         \label{fig04}
%   \end{figure}
   \end{figure}

\begin{figure*}
 %  \begin{figure}
   \centering
          \includegraphics[angle=90, width=0.94\textwidth]{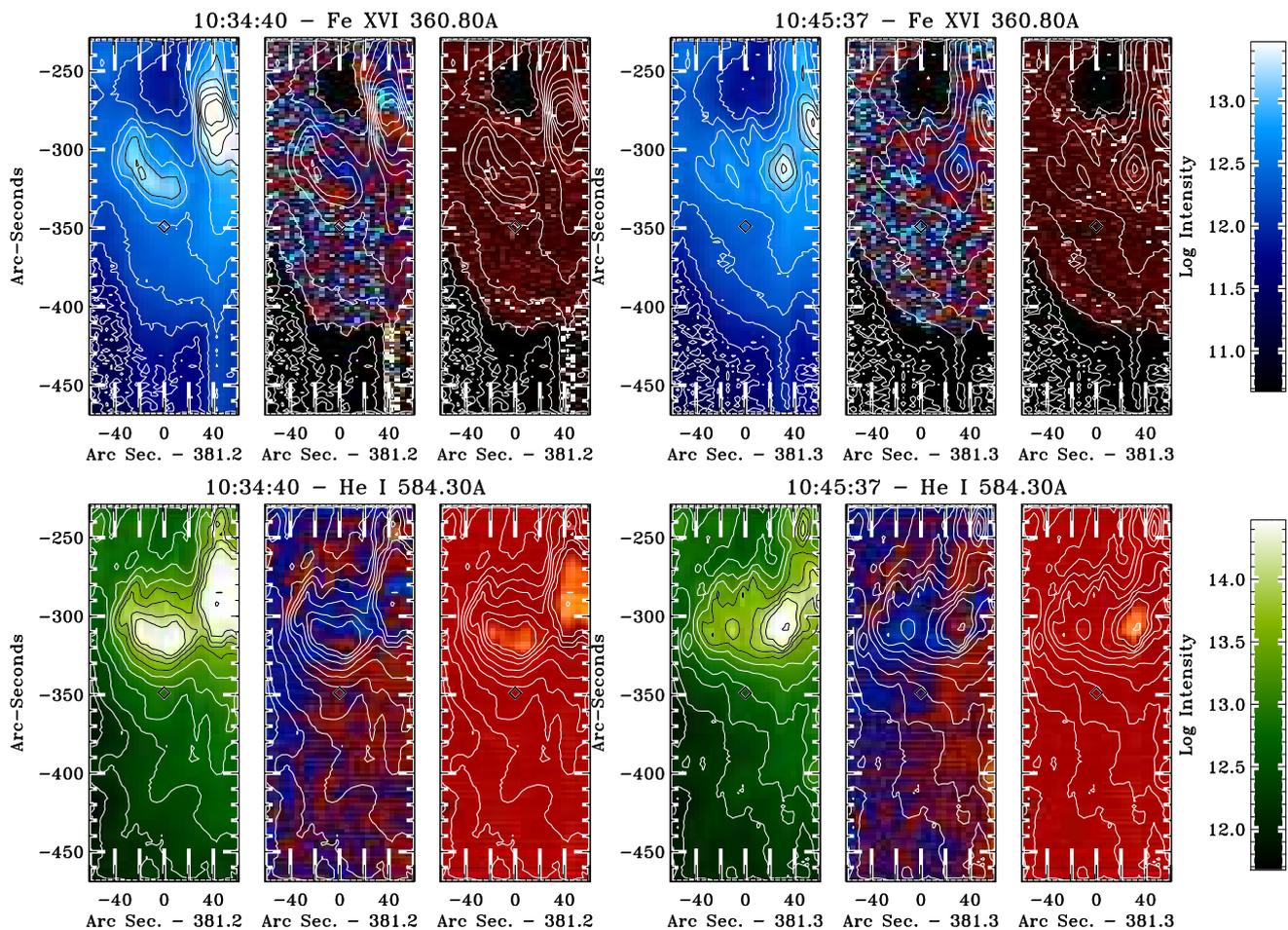}
      \caption{Similar to Fig.~\ref{fig03} but the CDS rasters are for \ion{Fe}{XVI} 360.80\,$\AA$ (top row) and \ion{He}{I} 584.30\,$\AA$ (bottom row) lines at 10:34:40\,UT and 10:45:37\,UT on 2002 August 19. Places in the \ion{Fe}{XVI} image that were below a certain intensity threshold have not been fitted. The velocity and FWHM panels use the same colour tables as Fig.~\ref{fig03}. 
              }
         \label{fig05}
 %  \end{figure}
   \end{figure*}

Fig.~\ref{fig04} depicts the broadened Gaussian fits to the \ion{O}{V} line intensity profiles for one of those pixels showing the strong line broadening at 10:34:40\,UT and 10:45:37\,UT. The quiet line profiles outside the flaring region are also shown. These have widths comparable to the instrumental width. Fig.~\ref{fig05} shows the CDS raster images for \ion{He}{I} 584.30\,$\AA$ and \ion{Fe}{XVI} 360.80\,$\AA$ lines at times when the strong \ion{O}{V} line broadenings are seen. Similarly to \ion{O}{V} the \ion{He}{I} line is fitted with a single broadened Gaussian function. Since there is a second weaker line on the blue side of the \ion{Fe}{XVI} peak (a blend of  \ion{Fe}{XIII} lines at 359.6\,$\AA$ and 359.8\,$\AA$) the \ion{Fe}{XVI} line is fitted with a double Gaussian where the FWHM is taken from the stronger central line. The corresponding line broadenings are seen in cooler \ion{He}{I} line, but not in the hot \ion{Fe}{XVI} line. The \ion{He}{I} line broadenings at 10:34:40 and 10:45:37\,UT are observed in the regions with the plasma down-flows. It is consistent with the observations in \ion{O}{V} line, though the plasma flows are more mixed in \ion{O}{V} line.

%-----------------------------------------------------------------

\begin{figure}
%   \begin{figure}
   \centering
      \includegraphics[width=0.48\textwidth]{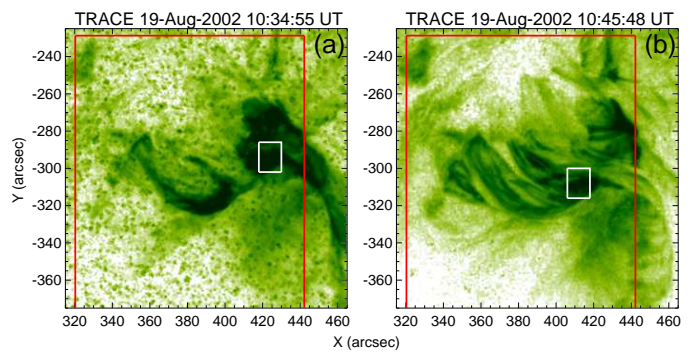}
      \caption{TRACE 195\,$\AA$ EUV direct images (with the reverse color scale where dark corresponds to high intensities) of the source M2.1 flare in event 2 around the flare maximum (panel a) and ten minutes later (panel b). Red rectangles depict the CDS FOV. White rectangles mark the pixels with the enhanced line broadenings. Note that panel (a) shows in a better resolution the EIT flare in Fig.~\ref{fig01}b. The noise spikes in the images are not filtered out.  
              }
         \label{fig06}
%   \end{figure}
   \end{figure}

The intensity brightening at 10:34:40\,UT occupies two separate areas. As indicated in the EIT image (5$\arcsec$ spatial resolution) in Fig.~\ref{fig01}b the brighter area on the west corresponds to the jet locus and the less bright region on the east to the top of the closed loop. The locus of the jet originated at the southwest boundary of a large sunspot at the interface with the minor polarity field. Note that the negative IMF polarity during the event 2 is consistent with sunspot polarity field. To see more details of the jet we plot in Fig.~\ref{fig06}  the high-resolution (1$\arcsec$) TRACE 195 EUV images at times of the enhanced line broadenings. The pixels with the line broadening at 10:34:40\,UT coincide with the jet locus and the area where the coronal closed structure reconnects with the open field lines. During the second line broadening at 10:45:37\,UT a flare/small-jet was observed shortly between 10:45--10:50\,UT with the TRACE high cadence (17 sec) EUV images. The post flare arcade of the closed loops is well seen in the upper-right part of the image in Fig.~\ref{fig06}b. 

To align between TRACE and SOHO CDS we at first check the inclinations of different flaring structures in SOHO EIT 195\,$\AA$ image on 2002 August 19 10:36:10\,UT (Fig.~\ref{fig01}b) and TRACE 195\,$\AA$ image on 2002 August 19 10:34:54\,UT (Fig.~\ref{fig06}a). Note that the time offset of 75 seconds between these two images corresponds to a negligible 0.1$\arcsec$ in the solar rotation. The alignment was done after rescaling TRACE to L1. It is obvious that no significant rotation is needed to co-align the jet seen in both images. In the further step the TRACE 195\,$\AA$ images in Fig.~\ref{fig06} were manually shifted in solar X by $+$10$\arcsec$ to co-align with features observed in CDS \ion{Fe}{XVI} 360.80\,$\AA$ line raster image (Fig.~\ref{fig05}). The similar approach was applied to the TRACE image in Fig.~\ref{fig01}c. The EIT 195\,$\AA$ channel includes emission from the bright \ion{Fe}{XII} line formed at $\sim$1.6\,MK and therefore the comparison with CDS 2.5\,MK \ion{Fe}{XVI} line is reasonable.

%-----------------------------------------------------------------

\subsection{X-ray and radio observations in event 2}

\begin{figure}
%   \begin{figure}
   \centering
      \includegraphics[width=0.48\textwidth]{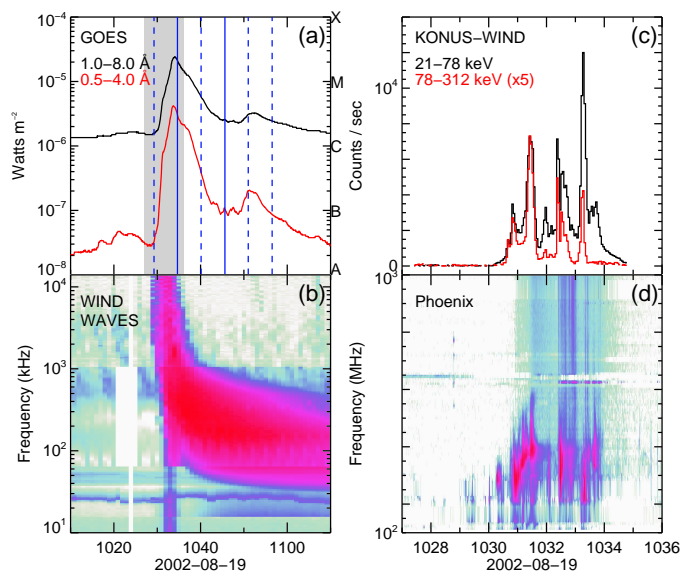}
      \caption{(a) Full-disk GOES-10 3-sec X-ray fluxes. The labels A, B, C, M, and X on right y-axis indicate flare classes in 1--8\,$\AA$ channel. The vertical lines mark the start times of six CDS rasters; two solid vertical lines indicate the starts of the rasters when the excess line broadening is observed. (b) Wind WAVES 1-min averaged radio spectra. (c) Full-disk Wind KONUS 3-sec averaged X-ray counts$^{-1}$ in a 9-minute period marked by a shaded bar in panel (a). The 78--312\,keV fluxes are multiplied by a factor of 5. (d) Phoenix 100\,ms radio spectra. 
              }
         \label{fig07}
%   \end{figure}
   \end{figure}

Fig.~\ref{fig07}(a--b) shows GOES X-ray flux (1.0--8.0\,$\AA$, 0.5--4.0\,$\AA$) light curves and Wind WAVES radio spectrogram during the M2.1 class flare in event 2. A strong type III radio burst was observed during the impulsive (or rapid X-ray flux increasing) phase of the flare. The line broadening observed at the raster 10:34:40\,UT (marked by a leftmost solid vertical line) corresponds to the time just after the flare maximum and coincides with the termination of the type III radio burst (or a termination of energetic electron injection into interplanetary space). More accurately, the pixels with the enhanced line broadening were scanned even slightly later at $\sim$10:35:10--10:35:40\,UT. The same area was scanned in the previous rastering, around the flare (and type III burst) onset time at $\sim$10:29:45--10:30:15\,UT, but not yet enhanced line broadening is seen. The excess line broadening at the raster 10:45:37\,UT (marked by another vertical solid line) most likely corresponds to another X-ray flare, not well resolved from a high background during the decay phase of M2.1 flare. We can see that the GOES X-ray flux in the M2.1 flare does not continue decreasing to the background values, but forms a plateau around 10:45\,UT that may correspond to the $<$C2.5 class flare. There were no type III radio bursts associated with this second line broadening. Either the jet might have formed along the closed magnetic field lines and/or the energy release was too weak for a particle acceleration. %A closed configuration is further supported by the broad red wing in the asymmetric line profile (see right panel of Fig.~\ref{fig04}) indicating the enhanced plasma down-flows.

After examining full-disk EIT difference images (12-min cadence) we note that the follow up C3.2 X-ray flare with a maximum at $\sim$10:52\,UT on 2002 August 19 (see Fig.~\ref{fig07}a) originated in the same active region. Again, no type III radio bursts were accompanied by this flare. The low energy electrons ($<$30\,keV), attributed to the source of type III radio bursts \citep{kru99}, have been measured in event 2 by Wind 3DP telescope \citep{lin95} in eight energy bins between 1.8 and 19\,keV. However, due to a large pre-event background, the onset of the low-energy electron event is not well measured.

%-----------------------------------------------------------------
%\fussy
Fig.~\ref{fig07}(c--d) shows high-resolution HXR and radio data in a 9-minute period, marked by a shaded bar in Fig.~\ref{fig07}(a--b), covering the impulsive and early decay phase of the M2.1 flare in event 2. The HXR intensity was detected by the Wind KONUS X-ray instrument in two energy channels $\sim$20--80 and $\sim$80--300\,keV. Multiple HXR bursts were detected between 10:30:40 and 10:34:00\,UT on August 19 (see also Wind KONUS Solar Flare database at \url{http://www.ioffe.ru/LEA/kwsun/}). The radio spectra, in the frequency range 0.1--1.0\,GHz, were observed with a Phoenix-2 Radio Telescope. A large group of intense type III bursts at 10:30:06--10:33:54\,UT coincides with HXR bursts indicating that their parent energetic electrons are related to the same acceleration episodes. The NOAA NGDC Solar Radio Bursts Report (\burl{ftp://ftp.ngdc.noaa.gov/STP/space-weather/solar-data/solar-features/solar-radio/radio-bursts/reports/spectral-listings/2002/}) lists a large group of decimetric (DCIM) bursts ($>$10) at 10:30:18--10:34:06\,UT in the frequency range 2.0--4.5\,GHz (Ond\v{r}ejov). Just above the range of the WAVES instrument the type III bursts continue in the range 20--100\,MHz observed with Artemis-IV (Thermopylae) radiospectrograph.

\section{Discussion and further prospects}

The spectral line broadening is usually caused by thermal ion motion. In solar flares the excess (non-thermal) line broadening has been often observed \citep[e.g.,][]{flu89}. It has been discussed that the excess broadening observed in EUV spectral lines may be an indicator of unresolved plasma flows \citep[e.g.,][]{mil11}, wave turbulence \citep[e.g.,][]{moo12} or accelerated non-thermal ion motions \citep[e.g.,][]{jef16}.

\citet{jef17} have reported that a velocity distribution emitting from the \ion{Fe}{XVI} ($Q/A=$15/56) line has a higher excess broadening than the \ion{Fe}{XXIII} ($Q/A$=22/56) velocity distribution. This would be consistent with the measured heavy-ion enhancement that increases with low $Q/A$ ratio in \element[ ][3]{He}-rich flares \citep{rea04,mas04} as well as with the mechanism of ion acceleration in reconnection exhausts \citep{dra09} or cyclotron resonance with Alfv\'en waves \citep{kum17} that predict more efficient acceleration for ions with low $Q/A$. Note that \ion{O}{V} and \ion{Fe}{XVI} have almost identical $Q/A$ and therefore should exhibit similar enhancement factor. However, no excess broadening was observed in \ion{Fe}{XVI} line in the examined event. \citet{jef17} have found that broadened line profiles may arise from non-thermal Fe if ions are locally accelerated on timescales $<$0.1\,s. Due to these stringent conditions, the authors suggest that more plausible explanation for the observed line profiles are non-Gaussian turbulent velocities that was supported by the observations of small redshifts. Similarly, we observe small redshift at sites with the excess line broadening that could drive plasma turbulence responsible for the \ion{He}{I} and \ion{O}{V} line profiles.

%We found the excess broadening only at relatively cooler lines, but not at the hot lines that may correspond to higher altitudes in the corona. However, we measure the population of energetic ions (including Fe) that escape, through the corona, into the interplanetary space. If a number (or energy content) of the escaping ions presents only a small fraction of the total amount of the energetic ion production it may not produce an observable effect at higher altitudes in the corona. The number of escaping ions can be either more or less than the number of trapped ions in the flare site \citep{ram93}. Note that the number of 10-keV escaping electrons is much smaller ($<$1\%) than precipitating down to the chromosphere \citep[][ and references therein]{kru11}. 

\citet{orr76} have shown that non-thermal neutral hydrogen \ion{H}{I} may become excited and radiate Ly$\alpha$. The authors pointed out that the effect can occur in the lines of other ions and atoms. \citet{pet90} have extended the calculations for a singly ionized helium (\ion{He}{II}). The authors give the charge-exchange cross sections for formations of \ion{He}{II} from $\alpha$-particle and of neutral \ion{He}{I} from \ion{He}{II}. One can find that a probability of double charge-exchange process to form \ion{He}{I} rapidly decreases with the energy increase in 50--400\,keV\,nucleon$^{-1}$, implying that the effect would not be strong in the \ion{He}{I} line; but details on a formation of broadened line profiles from the energetic \ion{He}{I} are unknown. 

It has been thought that ions in \element[ ][3]{He}-rich SEP events are accelerated at the impulsive phase of solar flares along with non-relativistic electrons \citep{rea85,wan12}; \citet{rea88b} has suggested that ions are accelerated at time of HXR maximum. This might also be the case of event 2 where the ion and (the relativistic) electron injection times roughly coincide. In the examined event, the strongest line broadening was seen near the reconnection site of the jet, however, at the time when the type III radio and HXR bursts terminated. This implies that the enhanced line broadening occurred at times when there was no energetic electron injection into interplanetary space or to the chromosphere. Unfortunately, this location was not scanned during the impulsive phase of the flare, though at the flare onset time no line broadenings were seen. Nevertheless, the observed excess in the line broadening appears not to be related to the wave turbulence that accelerated electrons (and presumably also the ions) otherwise energetic particle generation would continue. 

\citet{win89} has developed an acceleration mechanism related to the bulk flow evaporation, based on the ion-ion stream instability initiated by the precipitating energetic electrons. Though the observations in event 2 show a blueshift in the SEP source flare the strongest blueshift was observed in the weaker follow-up recurrent X-ray flare ($\sim$15 minutes after the SEP event associated X-ray flare maximum) without a signature of particle acceleration. \citet{rea94} have doubted the mechanism as the events with chromospheric evaporation did not correlate with \element[ ][3]{He}-rich flares. The question whether such mechanism could operate in \element[ ][3]{He}-rich SEP flares remains still open. But it appears that the intensity of the blueshifted component alone may not be an appropriate indicator for the mechanism at work.

Diagnostic capabilities of EUV line spectroscopy for ion acceleration in solar flares need to be further explored, and possibly with better temporal resolution observations. Note that in the examined event, the area around the jet locus was not scanned during the critical impulsive phase that lasted $\sim$3--4 minutes. Thus, we do not know when the excess broadening has started and how it evolved in time. For many flares, the excess broadening is observed to increase at the beginning of the impulsive phase \citep{flu89}. Furthermore, near the flare onset when the background from the flaring chromosphere is small and charge-exchange is more efficient, as the medium is less ionized, the effect from the non-thermal ions may be better visible \citep{orr76,pet90}. More recent spectroscopic observations are performed with EIS and IRIS spectrometers. The EIS, in operation since November 2006, observes the solar corona at two wavelength bands, 170--210\,$\AA$ and 250--290\,$\AA$ with 0.06\,$\AA$ spectral and 2$\arcsec$ spatial resolution over a FOV of up to 360$\arcsec$ $\times$ 512$\arcsec$ \citep{cul07}. A preliminary check of the EIS flare catalog (\url{http://solarb.mssl.ucl.ac.uk/SolarB/eisflarecat.jsp}) indicates 10 out of 54 (January 2007--May 2014) already reported \element[ ][3]{He}-rich SEP source flares \citep{nit15,buc16} in the EIS FOV. The IRIS provides UV spectra and images around 1400\,$\AA$ and 2800\,$\AA$ that focus on the chromosphere with 0.03\,$\AA$ spectral, 0.4$\arcsec$ spatial, and 2-second temporal resolution over a FOV of up to 175$\arcsec$ $\times$ 175$\arcsec$ \citep{dep14}. So far only nine \element[ ][3]{He}-rich flares in the Earth view have been reported \citep{nit15} that cover an observing period of IRIS (since July 2013) but no one is found in the spectrograph FOV. An intentional pointing of the spectrometers to the candidate \element[ ][3]{He}-rich SEPs source (like an active region on near-equatorial coronal hole boundary on the western hemisphere) might provide more events for a systematic investigation.
\\
\begin{acknowledgements}
This work was supported by the Deut\-sche For\-schungs\-ge\-mein\-schaft (DFG) under grant BU~3115/2--1 and Max-Planck-Gesellschaft zur F\"orderung der Wissenschaften. \v{S}M acknowledges VEGA grant 2/0155/18.
\end{acknowledgements}


\end{document}